# Triadic closure in two-mode networks: Redefining the global and local clustering coefficients

**Tore Opsahl**
Imperial College Business School, Imperial College London
(t.opsahl@imperial.ac.uk)

Draft date: May 6, 2011

**Abstract**
As the vast majority of network measures are defined for one-mode networks, two-mode networks often have to be projected onto one-mode networks to be analyzed. A number of issues arise in this transformation process, especially when analyzing ties among nodes' contacts. For example, the values attained by the global and local clustering coefficients on projected random two-mode networks deviate from the expected values in corresponding classical one-mode networks. Moreover, both the local clustering coefficient and constraint (structural holes) are inversely associated to nodes' two-mode degree. To overcome these issues, this paper proposes redefinitions of the clustering coefficients for two-mode networks.

**Acknowledgements**
I gratefully acknowledge the comments of participants at the Conference and Workshop on Two-Mode Social Network Analysis 2009 in Amsterdam, The Netherlands, and the 2010 Sunbelt Conference in Riva del Garda, Italy. I would also like to thank Filip Agneessens, Martin Everett, Gerard George, Anushka Patel, John Skvoretz, and Tom Snijders for their constructive comments on an earlier draft of this paper.

**Introduction**

Networks are representations of systems in which the elements (or nodes) are connected by ties (Wasserman and Faust, 1994). Most networks are defined as one-mode networks with one set of nodes that are similar to each other. However, several networks are in fact two-mode networks (also known as affiliation or bipartite networks; Borgatti and Everett, 1997; Latapy et al., 2008). These networks are a particular kind, with two different sets of nodes, and ties existing only between nodes belonging to different sets. A distinction is often made between the two node sets based on which set is considered more responsible for tie creation (primary or top node set) than the other (secondary or bottom node set).

One of the first two-mode datasets to be analyzed was the Davis' Southern Women dataset (Davis et al., 1941), which recorded the attendance of a group of women (primary node set) to a series of events (secondary node set). A woman would be linked to an event if she attended it. Another category of two-mode networks that has become popular in recent years is scientific collaboration networks (Newman, 2001). The two sets of nodes are scientists and papers, and a scientist is linked to a paper if she or he is listed as an author. As scientists generally decide whether or not they would like to work on a paper, they are often assumed to be the primary nodes. However, it is not always obvious which node set is the primary one, and in these cases, the research question guides the choice. For example, in the case of interlocking directorates where the two node sets are directors and corporate boards, and ties represent affiliation of directors with boards, it is not clear whether directors or boards are the primary node set (e.g., Levine, 1979; Mizruchi, 1996; Seierstad and Opsahl, 2011). This is likely to be due to tie formation being a mutual process where the directors must (1) be invited to join the board, and (2) accept the invitation.

Two-mode networks are rarely analyzed without transforming them. This is because most network measures are solely defined for one-mode networks, and only a few of them have been redefined for two-mode networks (Borgatti and Everett, 1997; Latapy et al., 2008). Transforming a two-mode network to a one-mode network is often done using a method known as projection. This method operates by selecting one of the two node sets (often the primary node set) and linking nodes from that set if they were connected to at least one common node in the other set. Although the two-mode structure is discarded in this process, it is possible to define tie weights based on it. Specifically, the tie weights are often defined as the number of common nodes. This method was extended by Newman (2001) who argued that tie weights among authors in scientific collaboration networks should be discounted if the authors collaborated on papers with many others.[1]

The projection of two-mode networks creates a number of issues. First, each tie in a prototypical one-mode network is assumed to be created separately; however, this is not the case in projected two-mode networks. For example, while a standard phone call creates a communication tie from one person to another, a director forms ties with all the other directors on a board when she or he joins that board. This has direct implications for frameworks that utilize random networks to detect a baseline level (e.g., Opsahl et al., 2008) and when comparing measures observed in a network with those found in corresponding random networks. This is due to the fact that ties in classical random networks are assumed to be independent of each other (Erdos and Rényi, 1959).[2] Although this is neither the case in prototypical one-mode nor projected two-mode networks, the random networks are less comparable to projected two-mode networks than to prototypical one-mode networks as multiple ties can be created due to a single event in these networks. Second, depending on the degree distribution of the non-projected node set, a projected two-mode network tends to have more and larger fully-connected cliques than prototypical one-mode networks (Wasserman and Faust, 1994). These are produced when three or more nodes are connected to a common node in the two-mode network (e.g., all the directors on a single board are connected and form a fully connected clique). This feature impacts a number of network measures, especially those based on triangles including the structural holes measures (Burt, 1992, 2005) and the clustering coefficients (for a review, see Opsahl and Panzarasa, 2009). To exemplify the cliques, and the many triangles, produced when projecting a two-mode network, Figure 1 shows the main component of the interpersonal network among Norwegian directors (Seierstad and Opsahl, 2011).

---

[1] Formally, Newman (2001) defined the tie weights as $\sum_p \frac{1}{N_p - 1}$ where $p$ is the co-authored papers and $N_p$ is the number of authors on the paper $p$. 1 is subtracted from $N_p$ to ensure that the sum of tie weights are equal to the number of co-authored papers.

[2] Ties are independent of each other in classical random networks where the present of each tie is given by a fixed probability. In corresponding classical random networks, this probability is the observed density (i.e., $\frac{E}{N \times (N-1)/2}$ for an undirected network where $E$ is the number of ties, and $N$ is the number of nodes; Borgatti and Everett, 1997).

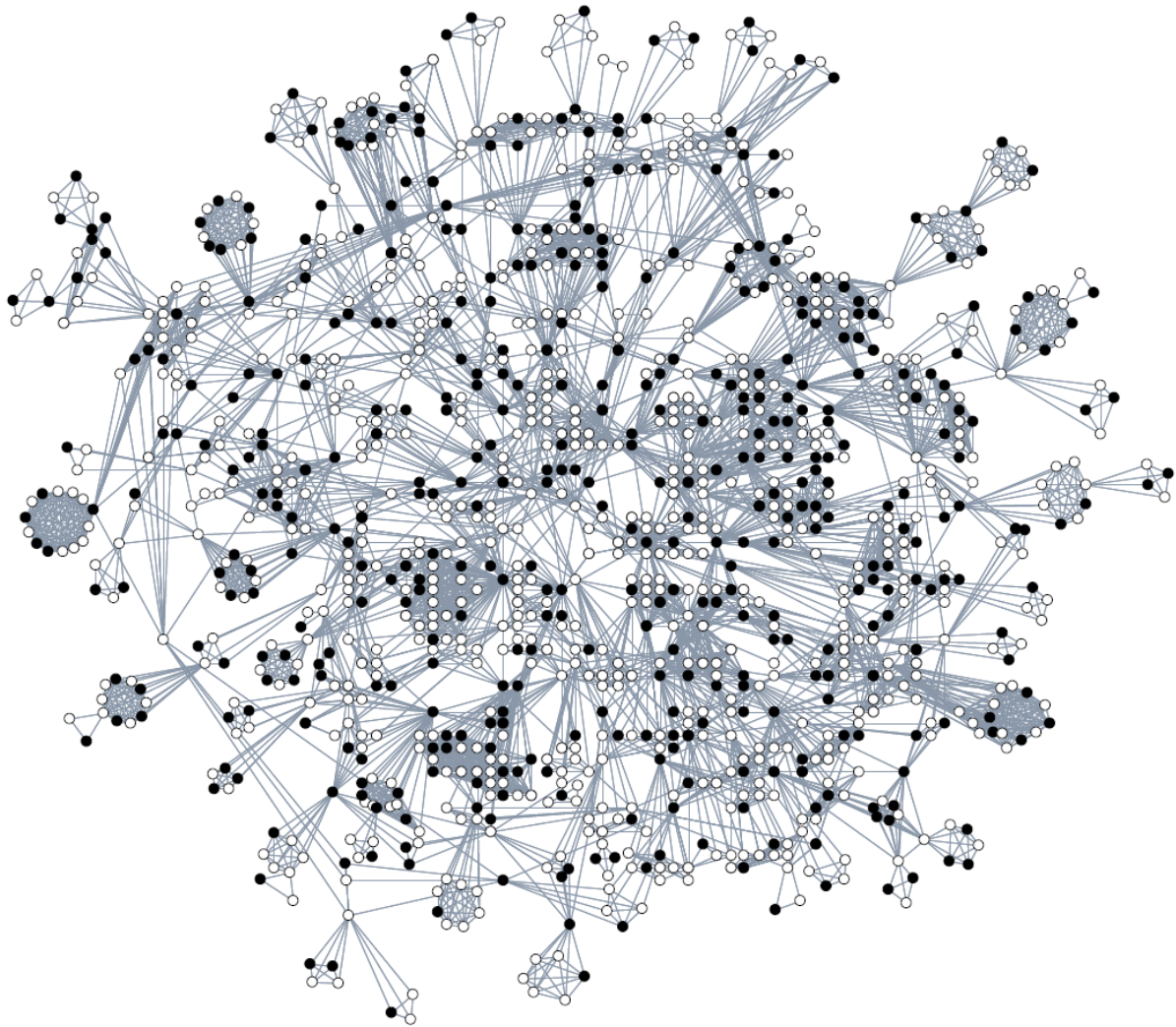

Figure 1: The network structure among directors (circles) who form part of the largest group of interconnected directors. Two directors are connected if they are members of the same board. The solid circles refer to women, whereas the hollow circles refer to men.

The rest of this paper is organized as follows. First, I will review the global and local clustering coefficients for one-mode networks and their properties. Then, I will explore which two-mode configurations create triangles when projected to a one-mode network. I will then propose novel measures that detect triadic closure in two-mode networks while omitting triangles formed by three nodes solely being connected to the same other nodes (e.g., three scientists writing a paper). For each of the proposed coefficients, I will suggest generalizations of the proposed measures to weighted two-mode networks. This will be followed by applications of the proposed measures to two-mode networks from the domains of event attendance, scientific collaboration, interlocking directorates, and online communication. Finally, I will offer some concluding remarks.

**Clustering coefficients for one-mode networks**

A subject that has long received attention in both theoretical and empirical research is nodes' tendency to cluster together. Evidence suggests that in most real-world networks, and especially in social networks, nodes cluster into densely connected groups (Holland and Leinhardt, 1970; Opsahl and Panzarasa, 2009). A number of measures have been developed for testing this tendency. Specifically, the *global* clustering coefficient assesses the overall level of clustering in a network (Luce and Perry, 1949), and the *local* clustering coefficient assesses the clustering in a single node's immediate network (i.e., the node and its contacts; Watts and Strogatz, 1998). This paper is concerned with both of these measures.

The global coefficient is the fraction of triplets or 2-paths (i.e., three nodes connected by two ties) that are closed by the presence of a tie between the first and the third node. It is formally defined as:

$$C = \frac{3 \times triangles}{triplets} = \frac{closed\ triplets}{triplets} = \frac{\tau_\Delta}{\tau} \quad (1)$$

where $\tau$ is the number of 2-paths, and $\tau_\Delta$ is the number of these 2-paths that are closed by being part of a triangle. This coefficient varies between 0 and 1. It is equal to 0 if no triangles exist in the network, and equal to 1 if all 2-paths are closed. In a completely connected network, the coefficient is 1 as all 2-paths are closed. Moreover, the value of the global clustering coefficient in a classical random network is equal to the probability of a tie being formed (i.e., the density) as ties are independent of each other (Erdos and Rényi, 1959). The global clustering coefficient has been generalized to weighted one-mode networks by Opsahl and Panzarasa (2009).

The local clustering coefficient measures the density in a node's local or ego network, and is the fraction of ties among a node's contacts over the possible number of ties between them (Watts and Strogatz, 1998). This coefficient is also linked to the concept of triplets or 2-paths as the denominator is equal to the number of 2-paths centered on a node, and the numerator is equal to the number of these where a tie is present between the first and third nodes. The local clustering coefficient for a node *i* can be formalized as follows:

$$C(i) = \frac{number\ of\ actual\ ties\ among\ node\ i's\ contacts}{number\ of\ possible\ ties\ among\ node\ i's\ contacts} = \frac{\tau_{i,\Delta}}{\tau_i} \quad (2)$$

where $\tau_i$ is the number of 2-paths centered on node *i*, and $\tau_{i,\Delta}$ is the number of these that are closed by being part of a triangle. While the global clustering coefficient is an aggregation of all 2-paths, the local one can be seen as an intermediary level of aggregation. The local clustering coefficient shares the same parameters as the global one, and has been generalized to weighted one-mode networks by Barrat et al. (2004).

Applying the two clustering coefficients directly to a two-mode network is senseless as two-mode networks are bipartite and, thus, the contacts of a node cannot be connected to each other by construction and no triangles can exist (Borgatti and Everett, 1997). Therefore, new measures must be devised. As generalizations and redefined measures should aim to maintain the purpose of the original measures, the following section will explore triadic closure and outline some of the research conducted in this area.

**Origins of triangles**

A key concept behind the clustering coefficients is triadic closure. Triadic closure is the addition of a tie that closes a 2-path to make it part of a triangle (e.g., the dashed line in Figure 2a). In a social setting, it can occur by a person introducing two contacts to each other or a person befriending friends' friends.

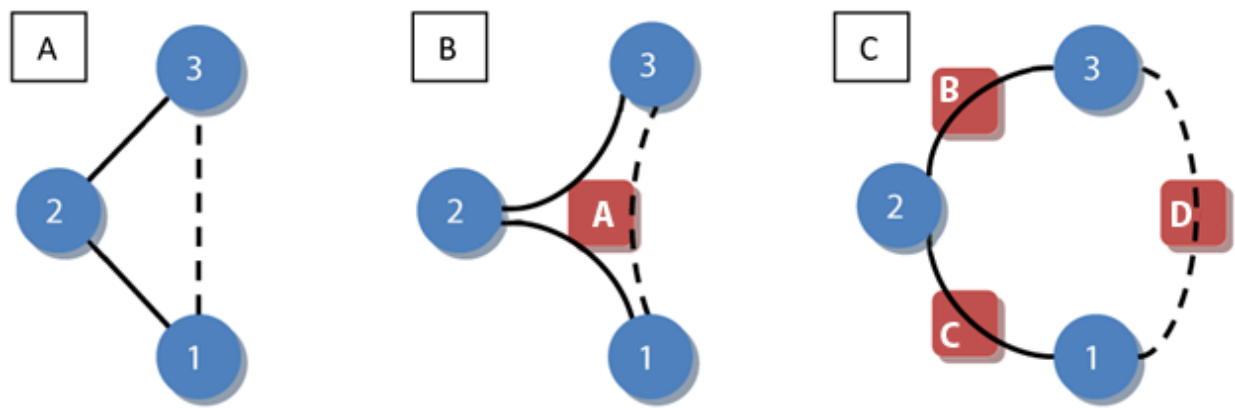

Figure 2: (A) A 2-path in a one-mode network (solid lines) that are closed by the first node forming a tie with the last node (dashed line), and thereby is a part of a triangle. (B-C) Two-mode configurations that become triangles if projected to a one-mode network.

Simmel (1923[1950]) was one of the first to reason about triadic closure. He argued that a strong social tie could not exist without being part of a triangle. In other words, a person is likely to share contacts with hers or his close contacts, and close contacts have an increased likelihood of knowing each other. This argument can be explored by comparing the weighted one-mode clustering coefficient with the binary one. Opsahl and Panzarasa (2009) showed that stronger 2-paths were more likely to be closed than weak ones in nine social networks and two non-social networks.

Granovetter (1973) built on Simmel's argument when he formulated the Strength of Weak Ties-theory. As close contacts have a higher likelihood of being connected than the acquaintances, they are more likely to move in the same social circles as the person and each other. In turn, their knowledge is more likely to overlap with the person's existing knowledge and each others'. Conversely, acquaintances or weak contacts are more likely to move in different social circles as they are further from the person. As such, the weak ties are likely to bring more novel information to the person.

The competitive angle of triadic closure was developed by Burt (1992, 2005). He argued that individuals occupied favorable positions if they had brokering opportunities among their contacts. While close-knit groups are cohesive with shared language and social control (Coleman, 1988), few opportunities exist for arbitrage or controlling the flow of information. Conversely, if contacts are themselves not connected (i.e., structural holes exist), then the individual that bridges them together is in a position to extract value.

A common factor for this line of research is analysis of ties among a person's contacts. In prototypical one-mode networks, this concept is fairly straight forward as ties are assumed to be created separately. Conversely, the same is not true in projections of two-mode networks as a single action might create new ties to contacts as well as creating ties among them. As illustrated by Figure 2b and 2c, a triangle in a projected two-mode network can be formed by two possible configurations. In Figure 2b, three primary nodes are connected to a common node, node A. Since this node creates the 2-path and closes it as well, all these 2-paths are closed by definition. On the contrary, the three primary nodes shown in Figure 2c become part of a 2-path when projected, but this 2-path is not closed by definition. This configuration would be closed if the first and last nodes of the path are connected to at least one common node, excluding the nodes on the path (e.g., node D). For this case, it can be argued that triadic closure occurs as a 2-path is formed from node 1 via nodes B, 2, and C to node 3, and is later closed by ties to node D.

**Global clustering coefficient for two-mode networks**
While the global clustering coefficient cannot be applied natively to two-mode data, it can be applied to one-mode projections of two-mode data. However, the clustering coefficient in one-mode classical random networks greatly underestimates the baseline-level of clustering in projected two-mode networks. In other words, the projected networks contain many more triangles than prototypical networks with a similar tendency for triadic closure. To illustrate this fact, I randomized the two-mode structure of a scientific collaboration network (see Empirical Test-section for details; Newman, 2001) while maintaining the degree distributions (i.e., randomly assigning the ties in the two-mode network while keeping each author's number of co-authored papers, and each paper's number of authors) before projecting it onto a one-mode network and calculating the global clustering coefficient. By using this randomization procedure, the triangles formed due to the two-mode structure are maintained in the one-mode projection. Across 1,000 projected random two-mode networks, the average global clustering coefficient was 0.1236, which is over 350 times larger than the coefficient in corresponding one-mode classical random networks.

To overcome this bias, a number of clustering coefficients for two-mode networks have been proposed in the literature (Lind et al., 2005; Latapy et al., 2008; Robins and Alexander, 2004; Zhang et al., 2008). These measures focus on 4-cycles, which is the smallest possible cycle in two-mode networks. For example, Robins and Alexander (2004) defined a coefficient as the ratio between the number of 4-cycles and the number of 3-paths. I have knowingly chosen not to follow this line of research. This is due to the fact that a 3-path would simply be, in the case of scientific collaboration networks, the number of papers that an author's collaborators have co-authored, and a 4-cycle would be two authors collaborating twice. Although this could be viewed as a form of clustering, it would not be triadic closure as it includes only two individuals, which is the topic of this paper. In fact, it could be considered a measure of reinforcement between two individuals rather than clustering of a group of individuals. From an evolving network perspective (e.g., in ERG models and SIENA; Peng et al., 2009; Snijders et al., 2011), it could be conceptualized as the adaptation or agreement that leads a node to connect to the same group as others.

The fundamental purpose of the one-mode clustering coefficient was to detect closure among three nodes. Based on this concept, I propose a new coefficient for two-mode networks that measures closure among three nodes from the primary node set instead of only two primary nodes (e.g., Robins and Alexander, 2004). Specifically, the denominator and numerator of the one-mode global clustering coefficient can be redefined in terms of 4-paths and closed 4-paths, respectively. This is due to the fact that all 4-paths in a two-mode network are 2-paths in a one-mode projection of the network; however, not all 2-paths in a one-mode projection are created from 4-paths. In fact, 2-paths can also be created due to multiple nodes being connected to the same node (i.e., the configuration in Figure2b). The 2-paths created by the latter mechanism would be excluded when only considering 4-paths in the two-mode structure. This feature is illustrated in Figure 3a and 3b. In the first panel (A), there are five 4-paths, three of which are closed.[3] These 4-paths represent five 2-paths in the one-mode projection (panel B). However, in the one-mode projection, there are an additional three 2-paths.[4] These are created among node 2, node 3, and node 4 as these nodes are all connected to node C in the two-mode network. The clustering coefficient of the two-mode network (panel A) is 0.6, while the clustering coefficient of the one-mode projection (panel B) is 0.75.

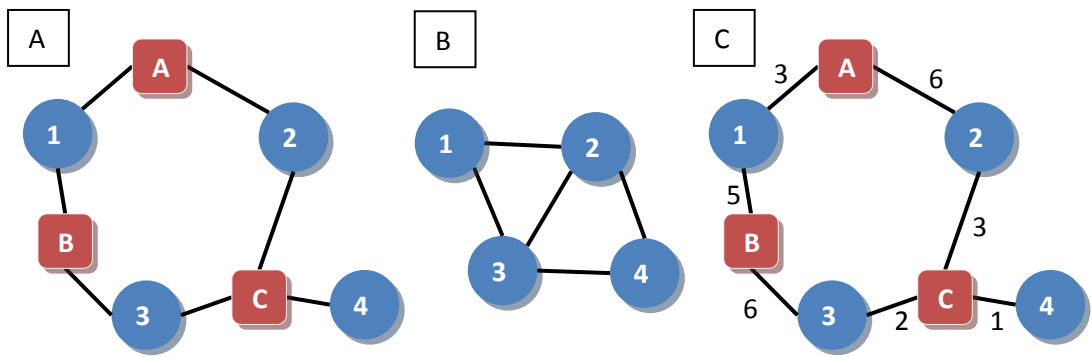

Figure 3: (a) A two-mode network where the shape and color of nodes represent the node set to which a node belongs, and (b) the one-mode projection of the round blue nodes in the two-mode network in panel A. The round blue nodes are the primary nodes in this projection. (c) A weighted two-mode network with a similar topology as the binary two-mode network in panel A.

---

[3] The 4-paths are 1-A-2-C-3 (closed by node B); 1-A-2-C-4; 1-B-3-C-2 (closed by node A); 1-B-3-C-4; 2-A-1-B-3 (closed by node C).

[4] These 2-paths are: 1-2-3 (closed); 1-2-4; 1-3-2 (closed); 1-3-4; 2-1-3 (closed); 2-3-4 (closed); 2-4-3 (closed); 3-2-4 (closed).

Formally, the proposed coefficient can be defined as:

$$C^* = \frac{closed\ 4paths}{4paths} = \frac{\tau^*_\Delta}{\tau^*} \quad (3)$$

where $\tau^*$ is the number of 4-paths, and $\tau^*_\Delta$ is the number of these 4-paths that are closed by being part of at least one 6-cycle (i.e., a loop of six ties with five nodes).

The coefficient has a number of properties. First, it varies between 0 and 1 as the numerator and denominator are positive numbers, and the numerator is a subset of the denominator. Second, in a fully connected network, it is equal to 1 as all 4-paths are closed. Third, in classical random two-mode networks (i.e., with a set number of nodes and density), the expected value is $1 - (1 - d^2)^{(N_p - 2)}$, where $N_p$ is the number of secondary nodes and $d$ is the density (Borgatti and Everett, 1997).[5]

The one-mode global clustering coefficient has been extended to weighted networks by assigning a value to each triplet or 2-path, $\omega$ (Opsahl and Panzarasa, 2009). This value is based on the weights of the two ties that compose the 2-path (e.g., the arithmetic mean, geometric mean, maximum, or minimum of the tie weights). The coefficient incorporates these values by being defined as the total value of closed 2-paths over the total value of all 2-paths. In addition to having the same parameters as the original coefficient, this coefficient produces the same outcome as the original one if all 2-paths have the same value (e.g., 1 in a binary network) or if the weights are randomly reshuffled in the network. In a similar spirit, the global coefficient proposed in this paper can be generalized to weighted two-mode networks, such as those created from online forums where the weights are the number of messages or characters posted to a thread. Specifically, a 4-path-value could be defined instead of a 2-path value. This value should be constructed based on the four tie weights, and could be defined using the same four methods as the one-mode weighted clustering coefficient. For example, the 4-path from node 1 to node 4 (via nodes A, 2, and C) in Figure 3c would be assigned a value of 3.25, 2.71, 6, or 1 if the arithmetic mean, geometric mean, maximum, or minimum was used, respectively. Formally, the coefficient for weighted two-mode networks could be defined as:

$$C^{*\omega} = \frac{total\ value\ of\ closed\ 4paths}{total\ value\ of\ 4paths} = \frac{\sum_{\tau^*_\Delta} \omega}{\sum_{\tau^*} \omega} \quad (4)$$

---

[5] The components of this express are the following. The density of the two-mode network, $d$, is the likelihood that a tie is present (i.e., $\frac{E}{N_i \times N_p}$ where $N_i$ is the number of primary nodes). The square of $d$ is the likelihood that (1) a tie is present from the first node on a 4-path to a node, and (2) a tie is present from that node to the last node on the 4-path. Thus, the inner bracket, $(1 - d^2)$, is the likelihood that these two ties are *not* present. By exponentiating the inner bracket with $N_p - 2$, it becomes the likelihood that *no* node connects the first and last node on the 4-path. The exponent is $N_p - 2$ as the two nodes on the 4-path cannot close it. By subtracting the likelihood that no node connects the first and the last node from 1, the outcome is the likelihood that a 4-path is closed if ties are randomly assigned in the network. Simulations conducted on ensembles of random networks with different number of nodes and ties produced outcomes that were not statistically different from the outcome attained with the above expression. Each ensemble contained 1,000 random networks.

The global clustering coefficients based on the four methods for the sample network shown in Figure 3c would be 0.6494, 0.6806, 0.6, and 0.7778, respectively. Given that all 4-paths contain a tie weight of 6 and the topology of the networks is identical to the binary two-mode network in Figure 3a, the coefficient attained with the maximum method is equal to the binary coefficient. The increases in the coefficients, when other methods for defining 4-path values are used, are a reflection of the fact that the closed 4-paths have relatively stronger ties than the open 4-paths. The various explanations given in Opsahl and Panzarasa (2009) for the differences between the methods for defining 2-paths values are also applicable in the case of 4-path values.

The generalized coefficient has a number of properties. In addition to having the same properties as the binary two-mode clustering coefficient, the generalized coefficient is equal to the binary one when all ties have the same value (i.e., binary). Moreover, the outcome of the generalized coefficient is approximately equal to the binary one if tie weights are randomly reshuffled in the network.[6]

**Local clustering coefficient for two-mode networks**

The local global clustering coefficient suffers from the same limitation as the global clustering coefficient, in that it cannot be applied directly to two-mode data, and a number of issues exist if applied to projections of two-mode data. Specifically, it is inversely related to nodes' two-mode degree (i.e., the number of ties in the two-mode network). If a node is connected to a single node in the two-mode network with at least two others, it will automatically have a clustering coefficient of 1 as the node's contacts are connected by default. To highlight the relationship between the local clustering coefficient and nodes' two-mode degree, Figure 4a shows the local clustering coefficient for nodes in Newman's (2001) scientific collaboration network where the ties have been randomly allocated in such a way that each node (i.e., authors and papers) maintain their number of ties. As can be seen, in random versions of a sparse two-mode network, the local clustering coefficient is roughly the inverse of nodes' two-mode degree, and not simply the density of ties. To emphasize that this limitation also affects the closely related structural holes measures (Burt, 1992), the analysis is repeated for the constraint-measure in Figure 4b.

---

[6] This result is based on simulations of 1,000 random networks with the same topology, but with the weights randomly reshuffled in the network (Opsahl et al., 2008). The binary clustering coefficient was not statistically different from the attained coefficients.

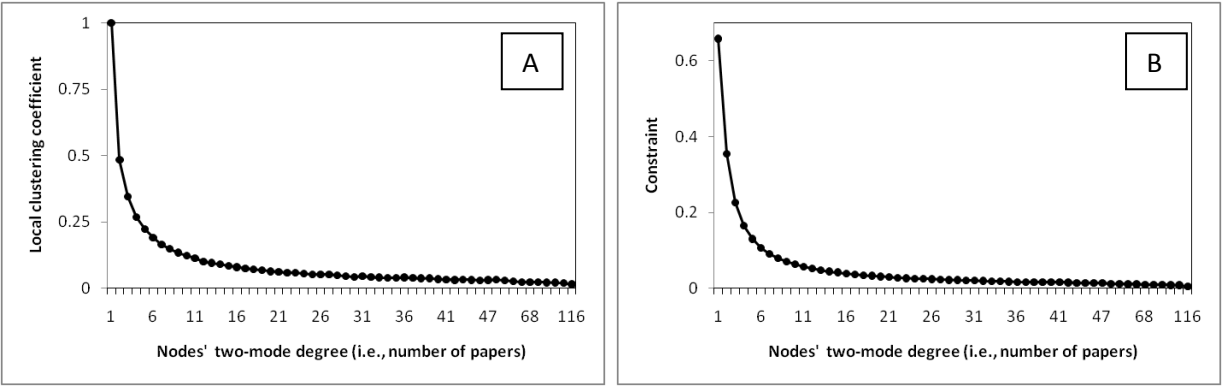

Figure 4: Association between primary nodes' two-mode degree and one-mode network measures computed on the projection of a randomly reshuffled sparse two-mode network where each node maintained the number of ties. For each level of degree, an average score is calculated among the nodes with that degree in 10 random versions of the network. (a) The local clustering coefficient. The values can be fitted by 1.02degree$^{-0.93}$ with an R² of 0.9881. (b) Burt's (1992) structural hole-measure constraint. The values can be fitted by 0.75degree$^{-1.07}$ with an R² of 0.9879.

The local clustering coefficient can be redefined in a similar vein as the proposed global clustering coefficient. While the one-mode local coefficient was based on 2-paths centered on a focal node, this could be extended to 4-paths centered on a focal node in two-mode networks. This would imply that the first and last nodes of the path are of the same node set as the focal node. For example, node 3 in Figure 3a is in the center of two 4-paths, where node 1 can be seen as the first node, and nodes 2 and 4 as the last ones. A 4-path does not exist from node 2 to node 4 (via node C, node 3, and node C) as node C is part of it twice. If such a path would be included in the measure, it would be closed by definition.

Formally, I propose:

$$C^*(i) = \frac{closed\ 4paths\ centered\ on\ node\ i}{4paths\ centered\ on\ node\ i} = \frac{\tau^*_{i,\Delta}}{\tau^*_i} \quad (5)$$

where $\tau^*_i$ is the number of 4-paths centered on the focal node $i$, and $\tau^*_{i,\Delta}$ is the subset of these in which the first and the last nodes of the path share a common node that is not part of the 4-path (i.e., part of at least one 6-cycle).

This coefficient has similar properties as the one-mode coefficient. First, for each node, the coefficient varies between 0 and 1 as the numerator and denominator are positive numbers, and the numerator is a subset of the denominator. Second, all 4-paths are closed in a fully connected network, and therefore, coefficient is equal to 1. Third, if ties are randomly placed in the network, the expected value of the local clustering coefficient is the same as the one for the global coefficient, $1 - (1 - d^2)^{(N_p - 2)}$.[7]

By using the generalization of the local clustering coefficient for weighted one-mode networks (Barrat et al., 2004), the proposed local clustering coefficient can be generalized for weighted two-mode networks. By using the same 4-path values as the global coefficient $\omega$, it is possible to differentiate them. In turn, it is possible to define a local clustering coefficient for weighted two-mode networks that share the same properties as the binary one, and is roughly equal to the binary one if weights are randomly assigned to ties. Formally, it could be defined as follows:

$$C^{*\omega}(i) = \frac{\text{total value of closed 4paths centered on node } i}{\text{total value of 4paths centered on node } i} = \frac{\sum_{\tau^*_{i,\Delta}} \omega}{\sum_{\tau^*_i} \omega} \quad (6)$$

**Empirical test**

To illustrate the proposed global clustering coefficient, I apply it to Davis's Southern Women dataset (Davis et al., 1941), a scientific collaboration network (Newman, 2001), the interpersonal network among directors on Norwegian public limited company boards (Seierstad and Opsahl, 2011), and data collected from an online forum. For this analysis, I did not rely on the expected values from classical random network, but used simulations to define a null distribution of values. In turn, the simulations allow for an analysis of whether observed coefficients are located in the extreme tails of the null distribution, and hence, significant.

First, Davis' Southern Women data were collected in the 1930s, and contains the attendance of 18 women to 14 events (Davis et al., 1940). This network is relatively dense as 91 percent of the possible ties are present and a clustering coefficient of 0.93 is found in the one-mode projection. In the simulations, 81 percent of networks exhibited a clustering coefficient lower than the observed. This implies that the observed value is not in the extreme tails of the null distribution, and suggests that it is inappropriate to argue that a triadic closure effect drives tie formation. Similarly, 44 percent of random networks had a clustering coefficient lower than the observed two-mode clustering coefficient (0.77). This further suggests that a triadic closure effect is not at play in this network. To highlight the relationship between the observed coefficients and the null distributions, Table 1 lists the coefficients and the 2.5 and 97.5 percentiles (i.e., the 95 percent confidence intervals) of the null distributions, and Figure 5 shows them graphically.

---Table 1 and Figure 5 about here---

Conversely, an above expected level of clustering exists in Newman's (2001) scientific collaboration network. This network is based upon 22,016 co-authored papers published on the arXiv e-repository website between 1995 and 1999. In total, these papers have 16,726 authors listed. The one-mode projection of this network, using the authors as the primary node set, has often been used as an empirical example for a variety of network measures (e.g., Opsahl et al., 2008). In the projected one-mode network, a clustering coefficient of 0.36 is obtained, and none of the corresponding random networks exhibited a clustering coefficient of this magnitude (the 97.5 percentile of the null distribution is 0.0004). Based on this result, it has been argued that there is an exceptionally strong tendency towards clustering (Newman, 2001). However, the one-mode clustering coefficient includes many of the triangles that are formed by construction as the average number of authors per paper is 2.66. The proposed clustering coefficient for two-mode networks can be applied in an effort to exclude these triangles. This coefficient is 0.28. None of the simulation networks showed a coefficient of this magnitude (the 97.5 percentile of the null distribution is 0.0006). Albeit a weaker effect than the one found in the projected one-mode network, this suggests that there is a strong triadic closure effect in this network.

---

[7] Simulations conducted on ensembles of random networks with different number of nodes and ties produced outcomes that were not statistically different from the outcome attained with this above expression. Each ensemble contained 1,000 random networks.

The Norwegian interlocking directorate contains a large number of cliques as shown in Figure 1. These cliques impact the one-mode clustering coefficients. This dataset is the interlocking directorate of Norwegian public limited companies on August 1, 2009 (Seierstad and Opsahl, 2011). Based on this dataset, I constructed the interpersonal network among the 1,495 directors, and linked two individuals if they sat on the same board. In this one-mode projection, the observed clustering coefficient is 0.68. None of the corresponding random networks showed a clustering coefficient of this magnitude (the 97.5 percentile of the null distribution is 0.0042). Conversely, in the two-mode network, the observed clustering coefficient is 0.0114. Although none of the simulated networks exhibit a clustering coefficient of that size (the 97.5 percentile of the null distribution is 0.0055), the observed two-mode value is substantially closer to the corresponding null distribution than the one-mode version. As such, a weaker triadic closure effect is observed than assumed from the projected one-mode network.

To illustrate the global clustering coefficient for weighted two-mode networks, I collected forum data from a Facebook-like online community of students at University of California, Irvine, in 2004. The overall community consisted of 2,995 students who could befriend each other and communicate using private messages. In addition, they could create groups and post broadcast messages to any group that they were a member of. In total, 889 students posted 33,720 broadcast messages to 552 groups. On average, each student posted 4.76 messages or 480.19 characters to each group they actively participated in. A two-mode network can be constructed from this dataset by linking a student to a group if she or he posted to it. Unlike the other datasets, it is possible to create both a binary and a weighted two-mode network from this dataset. The tie weights in the weighted two-mode network are based on the number of characters posted to groups by students. As can be seen from Table 1, the weighted coefficient is greater than the binary one. This suggests that stronger 4-paths are more likely to be closed than weaker ones. As such, this network confirms Simmel's (1923[1950]) assumption that a person is more likely to share contacts with close contacts also in two-mode networks.

To exemplify the proposed local clustering coefficient for two-mode networks, I used the Davis Southern Women dataset as the number of nodes is limited. Table 2 shows the local clustering coefficient scores attained from the two-mode network and projected one-mode network as well as the two-mode and one-mode degree scores (i.e., the number of events attended and the number of other women attending the same events, respectively). There are a number of observations. First, for all the nodes that did not have the maximum value, the two-mode coefficient is smaller than the coefficient attained on the projected network. This result is not automatic as multiple 4-paths might exist among three primary nodes. Therefore, the two-mode coefficient might be higher than the one attained on projected one-mode network. It gives a strong indication of the bias that is created by three or more primary nodes connected to a common node. Second, the difference between the two coefficients is greater for the women attending fewer events (pair-wise correlation between the difference in coefficients and the number of events is -0.69, with a *p*-value of less than 0.001). This corroborates the finding shown in Figure 4a suggesting that the bias is greater for nodes that attend fewer events.

---Table 2 about here---

To further highlight some of the benefits of redefining the local clustering coefficient, Figure 6 shows Flora and the network around her up to three steps. In the one-mode projection of the network, all the possible ties among Flora's contacts are present. This is due to the fact that eleven out of her twelve contacts attended event 9. The twelfth contact that did not attend event 9, Helen, is connected to all others through other events. The redefined clustering coefficient is less than 1 for Flora because event 9 and 11 are not used to forming closing ties among the women attending them (i.e., close 4-paths). Specifically, 4-paths exist from the nodes attached to event 11 to the nodes connected to event 9 (excluding themselves). In total, there are 31 4-paths, out of which 18 are closed by the events 6, 7, 8, and 10.

---Figure 6 about here---

**Conclusion**

Two-mode networks are rarely analyzed without transforming them into one-mode networks as there are only a few methods that can be directly applied to them (Borgatti and Everett, 1997). The main transformation method is called projection, and connects nodes from a chosen node set if they shared a common node. However, by projecting two-mode networks, certain assumptions in the one-mode methods might be violated, such as the ability of each tie to be formed separately. Moreover, measures based on triangles or ties among nodes' contacts might be biased. This is due to the fact that projected two-mode networks generally contain more and larger fully-connected cliques than regular one-mode networks. In particular, depending on the degree distribution of the non-projected nodes, the measure that assesses the overall level of clustering in a one-mode network, the global clustering coefficient, could be biased. More specifically, if the non-projected nodes have a degree greater than 2, triangles will be automatically formed in the one-mode projection, which will increase the coefficient. As shown in the analysis of a scientific collaboration network (Newman, 2001), the projection of a random version of the two-mode network has over 350 times the level of clustering that is expected in corresponding classical random one-mode networks. Thus, there is an acute need to redefine one-mode measures for two-mode networks, and in particular, the ones that are based on ties among nodes' contacts.

This paper proposed such a redefinition for both the global and local clustering coefficients. Both the redefinitions only consider the network structures that are not closed by definition (e.g., not three nodes that are connected to a common node in the two-mode structure). This allows for an assessment of triadic closure in two-mode networks that is not affected by the same modeling issues faced by the existing clustering coefficients on projected one-mode networks.

The proposed coefficients are not without their limitations. A major one is that one of the node sets must be considered to be responsible for tie generation, and designated as the primary node set. While the local clustering coefficient can be calculated for all nodes in a network by repeating the analysis twice, the same is not true for the global clustering coefficient. Primary nodes must be the first and last nodes of 4-paths. Although it is rare that the designation of primary nodes is not obvious, when this is the case (e.g., for interlocking directorates), the researcher might want to calculate the proportion of 4-paths starting from both node sets that are closed. Nevertheless, such a solution is highly subjective and should only be conducted if the node sets are strictly equal in the tie generation process. Moreover, the proposed local clustering coefficient is undefined for nodes with a two-mode degree less than two and a projected one-mode degree less than two. Although the one-mode coefficient is undefined for nodes with a one-mode degree less than two, the additional requirement of a two-mode degree less than two is likely to affect more nodes. For example, in scientific collaboration networks, a scientist must have co-authored at least two papers with distinct others to attain a score.

The proposed coefficients represent only a first step in the process of redefining one-mode network measures for two-mode networks. Although the clustering coefficients are particularly affected by the projection procedure, there are many other measures that are also affected, such as the array of structural holes measures (Burt, 1992). For example, constraint is inversely associated with nodes' two-mode degree in projections of random two-mode networks as shown in Figure 4b. Although the local clustering coefficient does not normalize the network time or energy of each node (i.e., divide tie weights on outgoing ties by the sum of outgoing tie weights so that the sum is equal to 1 for every node) nor incorporate the tie weight of the closing tie, it could act as a structural holes measure for two-mode networks. Nevertheless, specifically redefining these measures for two-mode networks could increase their accuracy, and might give rise to novel insights into organizing principles of two-mode networks.

|  | Davis Southern Women Club | Scientific Collaboration | Norwegian Directors | Online forum | |
| Network |  |  |  | Binary | Weighted |
|---|---|---|---|---|---|
| Primary nodes | 18 | 16,726 | 1,495 | 899 | … |
| Secondary nodes | 14 | 22,016 | 367 | 522 | … |
| Two-mode ties | 89 | 58,595 | 1,834 | 7,089 | … |
| One-mode ties | 139 | 47,594 | 4,065 | 71,380 | … |
| Observed one-mode GCC | 0.9284 | 0.3596 | 0.6805 | 0.5049 | … |
| One-mode GCC on classical random one-mode networks | 0.9088 [0.8750; 0.9399] | 0.0003 [0.0003; 0.0004] | 0.0036 [0.0031; 0.0042] | 0.1768 [0.1760; 0.1777] | … |
| Observed two-mode GCC | 0.7719 | 0.2769 | 0.0114 | 0.4954 | 0.5327 |
| Two-mode GCC on classical random two-mode networks | 0.7746 [0.6266; 0.8907] | 0.0006 [0.0005; 0.0006] | 0.0040 [0.0025; 0.0055] | 0.1117 [0.1061; 0.1171] | … |

Table 1: The global clustering coefficients (GCC) obtained on the four empirical datasets, and their random counterparts. The random values are the average coefficient found in 1,000 corresponding classical random networks, and the boundaries of the 95 percent confidence intervals are listed below. For the weighted two-mode clustering coefficient, 4-path values are the arithmetic mean of tie weights.

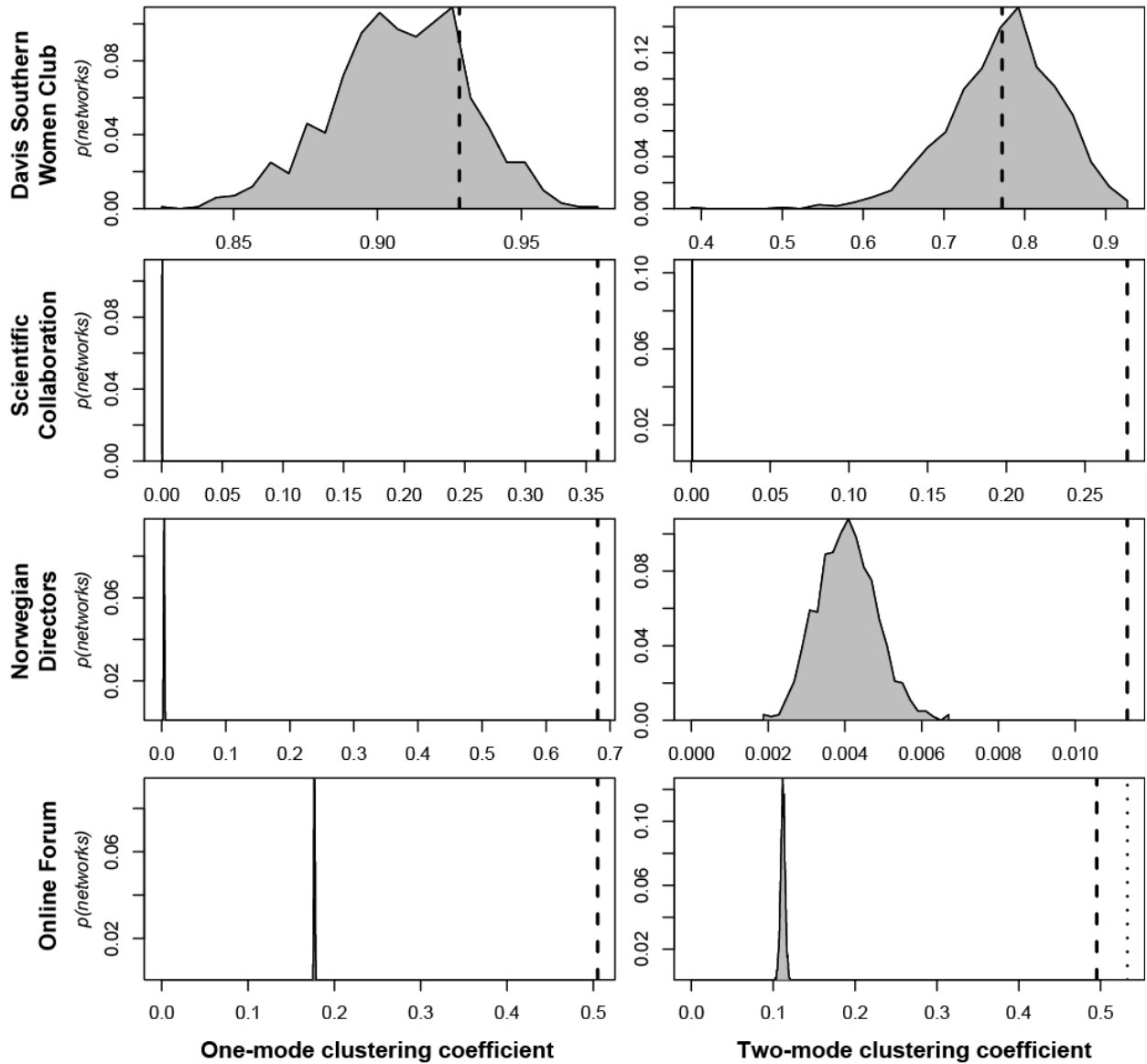

Figure 5: Graphical representation of the observed one-mode and two-mode clustering coefficients for the four empirical networks as well as their corresponding null distributions based on 1,000 classical random networks. Observed values are shown with vertical dashed lines (dotted line for the weighted two-mode clustering coefficient). The simulations results are shown by using 25 equally-spaced bins.

| Node | Events attended | Other women attending the same events | One-mode LCC | Two-mode LCC |
|---|---|---|---|---|
| EVELYN | 8 | 17 | 0.8971 | 0.7667 |
| LAURA | 7 | 15 | 0.9619 | 0.8422 |
| THERESA | 8 | 17 | 0.8971 | 0.7523 |
| BRENDA | 7 | 15 | 0.9619 | 0.8388 |
| CHARLOTTE | 4 | 11 | 1 | 1 |
| FRANCES | 4 | 15 | 0.9619 | 0.8690 |
| ELEANOR | 4 | 15 | 0.9619 | 0.7959 |
| PEARL | 3 | 16 | 0.9333 | 0.6463 |
| RUTH | 4 | 17 | 0.8971 | 0.6703 |
| VERNE | 4 | 17 | 0.8971 | 0.6741 |
| MYRNA | 4 | 16 | 0.9333 | 0.7139 |
| KATHERINE | 6 | 16 | 0.9333 | 0.7696 |
| SYLVIA | 7 | 17 | 0.8971 | 0.7462 |
| NORA | 8 | 17 | 0.8971 | 0.8380 |
| HELEN | 5 | 17 | 0.8971 | 0.8159 |
| DOROTHY | 2 | 16 | 0.9333 | 0.5407 |
| OLIVIA | 2 | 12 | 1 | 0.5806 |
| FLORA | 2 | 12 | 1 | 0.5806 |

Table 2: The two-mode and one-mode degree scores and the traditional and local clustering coefficients (LCC) of the women in Davis' (1940) Southern Women dataset. The randomly expected one-mode clustering coefficient is 0.9085, while the one for two-mode networks is 0.7978.

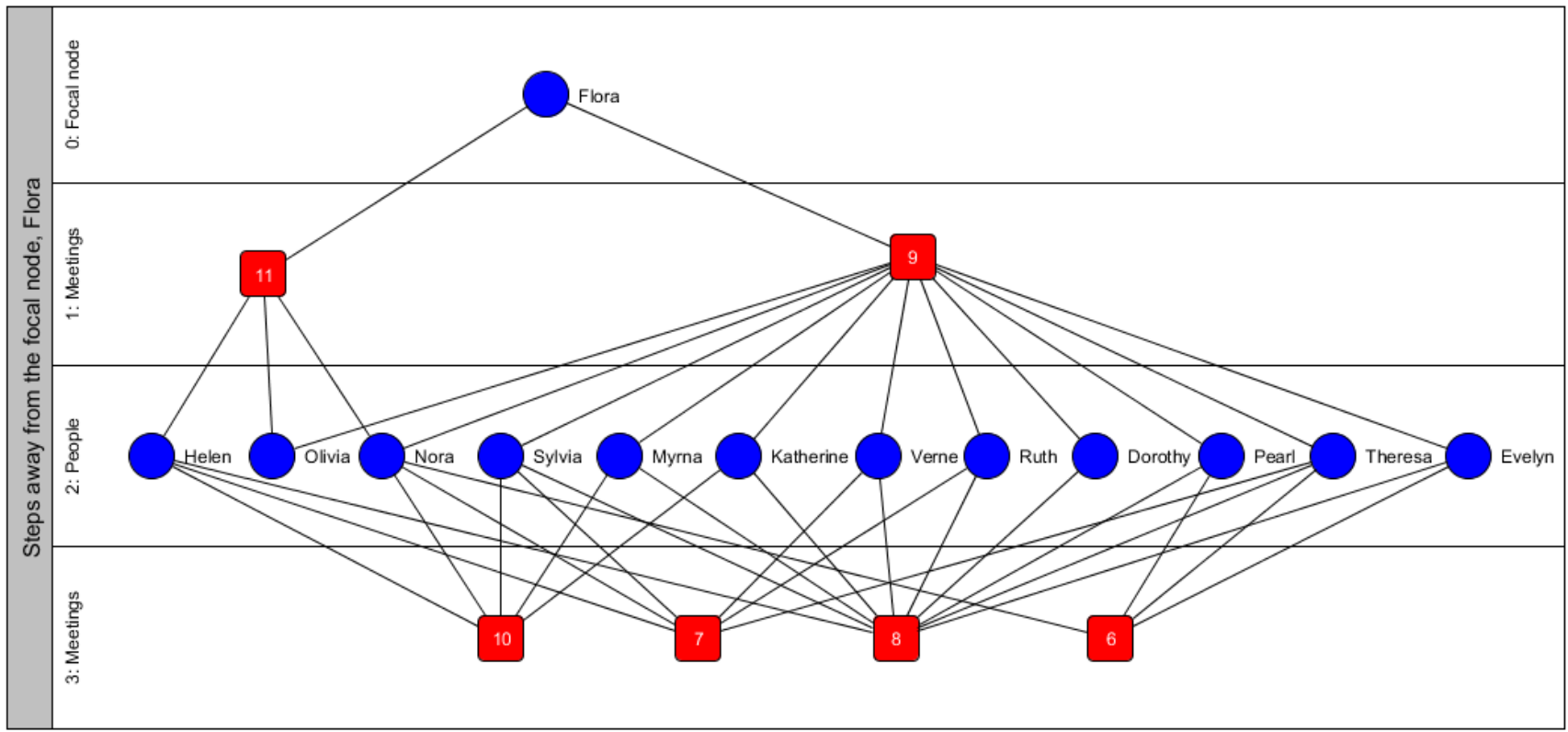

Figure 6: The local network (up to three steps) of Flora in Davis' (1940) Southern Women dataset. For clarity, only non-redundant nodes at the third step are shown.